\documentstyle[general_cite,psfig]{mn}

\title[A reanalysis of the X-ray luminosities of clusters of galaxies in the EMSS sample]{A reanalysis of the X-ray luminosities of clusters of galaxies in the EMSS sample with 0.3$<$z$<$0.6.}
\author[S.C. Ellis and L.R. Jones]
  {S.C.~Ellis\thanks{E-mail: sce@star.sr.bham.ac.uk} and L.R.~Jones \\
    School of Physics and 
Astronomy, University of Birmingham, Birmingham, B15 2TT, UK. \\}
\date{Accepted .....................; Received .....................;
in original form .......................} 

\begin{document}

\maketitle
\begin{abstract}

The X-ray luminosities of the \emph{Einstein} Extended Medium Sensitivity
Survey (EMSS) clusters of galaxies with redshifts 0.3$<z<$0.6 are remeasured 
using \emph{ROSAT} PSPC data.  It is found that the new luminosities are on 
average 1.18$\pm$0.08 times higher than previously measured but that this ratio
depends strongly on the X-ray core radii we measure.  For the clusters with 
small core radii, in general we confirm the EMSS luminosities, but for
clusters with core radii
$>$250 kpc (the constant value assumed in the EMSS), the new luminosities are 2.2$\pm$0.15
times the previous measurements. The X-ray luminosity 
function (XLF) at $0.3 < z < 0.6$ is recalculated and is found to be
consistent with the local XLF. The constraints on the updated properties of
the 0.3$<z<$0.6 EMSS sample, including a comparison with the
number of clusters predicted from local XLFs, indicate that the space density of luminous,
massive clusters has either not evolved or has increased by a small
factor $\sim$2 since $z=0.4$.
The implications of this result are discussed in terms of constraints on the
cosmological parameter $\Omega_{0}$.  

\end{abstract}

\begin{keywords}
galaxies:clusters:general --- X-rays:galaxies ---
cosmology:observations
\end{keywords}

\section{Introduction.}

The formation and growth of structure in the Universe depends upon certain
fundamental cosmological parameters which govern the environment in which
the structures form.  Clusters of galaxies, as the largest virialised
objects in the Universe, offer a unique insight into the formation of
structure and hence into the parameters governing their evolution.  

The X-ray properties of clusters offer a useful way of probing this
evolution.  The X-ray emission is due mainly to bremsstrahlung radiation
from the hot intra-cluster gas (which is the dominant component of baryonic
matter in clusters) and therefore the X-ray luminosity ($L_{\textrm{\scriptsize{X}}}$) is
expected to be positively correlated to the system mass.  A measurement of
the evolution of $L_{\textrm{\scriptsize{X}}}$ is therefore a natural choice for the study of the
evolution of clusters of galaxies.

Several surveys  have been made in the past which measure the evolution of
the X-ray luminosity function (XLF) with differing results.  Notably the
\emph{Einstein} Extended Medium Sensitivity Survey (EMSS) of \scite{hen92} and \scite{gio90} finds
moderate negative evolution of the number density of high redshift
clusters ($z>0.3$) of high $L_{\textrm{\scriptsize{X}}} > 5 \times 10^{44}$ ergs s$^{-1}$, i.e. there are more high
luminosity clusters in the
present than there were in the past, but no evolution of moderate
luminosity clusters.   This agrees well with simple
hierarchical formation theories in which the most massive clusters will
have formed most recently.  More recent
surveys, (\pcite{ros00}; \pcite{bur97}; \pcite{vik98a};
\pcite{jon98a}; \pcite{nic99}) find no evidence for evolution in the XLF at low to
moderate luminosities at $z<0.9$.  At high luminosities there remains some
ambiguity, probably due to the low numbers of high luminosity clusters in
the surveys.  There are three surveys, in addition to the EMSS, reporting
statistically significant evolution at high luminosities.  \scite{vik98a},
\scite{ros00} and \scite{gio01} all find moderate (typically a factor $\sim 
3$ in the space density) negative evolution of high luminosity clusters at
$z>0.3$ (or $z>0.5$ in the case of \pcite{ros00}).  \scite{nic99} also find
significant evolution but in a non-independent sample based partly on the
\scite{vik98a} survey.  Contrarily, in the survey of \scite{jon00} and
\scite{ebe01} no evidence for significant evolution is found at $z<0.9$.

Because any evolution of the XLF
would be most apparent at high luminosities (since more massive clusters 
evolve more quickly than less massive clusters in a bottom-up formation scenario),
and because evolution in this most critical part of the XLF is
still an area of controversy, it is thus desirable to extend observations of the evolution
of the XLF to high luminosity clusters at high redshift.  The EMSS remains
one of the only samples which contains a reasonable number of high luminosity,
massive clusters at high redshift and therefore it is the sample most
suited to constraining cluster evolution at high
$L_{\textrm{\scriptsize{X}}}$.

It has been suggested (\pcite{jon98a}) that the evolution of the XLF found
for high redshift clusters in the EMSS maybe partly an artefact due to an incorrect
conversion from detected flux to total flux.  Furthermore it is suggested
that a systematic increase in the fluxes of the EMSS $z>0.3$ clusters by a factor of
$\approx 1.25$ would account for the observed difference in the EMSS and
\emph{ROSAT} log$N$
--- log$S$ relations.  An explanation for such a difference
is offered by \scite{hen00} where it is pointed out that the inclusion of
the effect of the \emph{Einstein} point spread function (psf) was omitted
when correcting from detected flux to total flux in the original EMSS and
that the inclusion of such an effect would increase the total fluxes of
$z>0.3$ clusters by a mean
factor of 1.373.  However, it is also pointed out by \scite{hen00} that the 
inclusion of other effects, viz. integrating the cluster emission out only
as far as the virial radius (as opposed to infinity as in \pcite{hen92}) and
using a slightly different value for the mean King profile surface
brightness slope, $\beta$, would almost cancel out the correction due to
including the psf and thus the original EMSS formulation was fortuitously correct.

The corrections made in converting from detected flux to total flux in the EMSS were
large.  The original EMSS sample of X-ray clusters were detected using a $2.4'
\times 2.4'$ cell, and the flux falling outside this cell was corrected for
assuming a King profile with $\beta = \frac{2}{3}$ and a core radius of
$r_{\textrm{\scriptsize{c}}}=250$ kpc.  The average conversion factor from total flux to detected
flux is 1.8 for clusters at $z>0.3$ in the EMSS survey, and this 
factor is very sensitive to the assumed
core radius.  This correction was never claimed to be applicable to
individual clusters by \scite{hen92} but rather was used as a mean
correction for the sample.  There is some evidence that the core radius
assumed in deriving the correction is a good average (\pcite{vik98a},
Table~1 of this paper) although Fig~3. shows that for individual clusters
the measured core radius can have a significant effect on the derived luminosity.   

Therefore it is possible that the negative evolution seen in the EMSS
may not be representative of cluster evolution, and a reanalysis of the
luminosities of EMSS clusters is necessary to determine more accurately the
evolution of the XLF. 

\scite{nic97} have reanalyzed the EMSS sample using \emph{ROSAT} 
Position Sensitive Proportional Counter (PSPC) data
and find that there is evolution of the XLF, albeit at a lower rate than
originally found.  We feel, however, that a further reanalysis is 
justified for the reasons given in \S 2.1.

A reanalysis of the EMSS is also important because the sample has been used to
derive X-ray temperature functions (XTFs), which are used to derive
accurate values for $\Omega_{0}$.  Although the temperatures are
independent of the fluxes and core radii, which are remeasured in this paper, the volume
surveyed is not.  Therefore in order to compute the XTF accurately it is
necessary to have reliable measurements of the cluster fluxes.

In this paper we use a simple aperture photometry method to reanalyze the
fluxes of the EMSS sample for clusters with $z>0.3$ using \emph{ROSAT} PSPC 
data.  We use a large (3 Mpc radius) aperture so that the total fluxes are
almost independent of the model surface surface brightness profile.  The total fluxes are calculated firstly with the same King profile
used in \scite{hen92} for a direct comparison with the original
\emph{EINSTEIN} data.
The core radii of the clusters are then measured from surface brightness
fitting and the total fluxes are recalculated using the appropriately
corrected King profiles.  The XLF is then computed using the new data and
compared to the original XLF of \scite{hen92}.  It is assumed that
$H_{0}=50$ km s$^{-1}$ Mpc$^{-1}$ and $q_{0} = 0.5$ throughout.

\section{Data and analysis.}

Of the EMSS clusters listed in table 1 of \scite{gio94} (which
contains redshifts and luminosities of the EMSS clusters) 17 are at
$0.3<z<0.6$ and have been observed by \emph{ROSAT}. Of these, 11
have been observed with the PSPC .  Note that MS1333.3+1725 has now been
identified as a star (\pcite{lup99}, \pcite{mol99}), MS1209.0+3917 as a BL Lac
(\pcite{rec98}) and MS1610.4+6616 as (a point source at the position of) a
star (\pcite{sto99}).  There also exists a new redshift measurement for
MS1241.5+1710 of $z=0.549$ (\pcite{hen00}).

\subsection{Comparisons with the methodology of \protect\scite{nic97}.}

Although \scite{nic97} have reanalyzed the EMSS sample using \emph{ROSAT}
PSPC data we feel that a further reanalysis is necessary.  This is because
we feel that further improvements can be made to the reanalysis.

Our reanalysis differs from that of \scite{nic97} in three important
aspects.  Firstly, our samples of clusters at $z>0.3$ are different.  We
adopted the methodology of using the original sample of \scite{hen92} as
closely as possible but supplemented with newer measurements and
identifications wherever appropriate.  Accordingly we have used the revised
sample of \scite{gio94} throughout, except where recent measurements have
shown certain objects (viz. MS1333.3+1725, MS1209.0+3917, MS1610.4+6616;
see \S2) not to be clusters, and we have used new redshift measurements
where appropriate.  The methodology of \scite{nic97} is somewhat different
in that a measurement of the extent of the X-ray emission of each cluster
is made to decide whether the object is a cluster or not.  Sources which
are unresolved in \emph{ROSAT} PSPC data are flagged as uncertain.  Some of 
these sources were then examined in the HRI to check their extent, and for
those clusters for which there were no high-quality HRI data Monte Carlo
simulations were performed to assess the probability that such compact
sources would be observed by the PSPC.  Two clusters (at $z>0.3$) were
excluded by \scite{nic97} following this method, and two more were not excluded but
were classified as ambiguous.  This method of identifying clusters is not
satisfactory since MS1208.7+3928, which was excluded, has been found to be a
cluster (\pcite{sto99}); it is also included in the survey of
\scite{lup99} and is found independently in the survey of
\scite{vik98b}.  Also two clusters (MS1333.3+1725 and
MS1610.4+6616, see \S2) which are included by \scite{nic97} have since been
found not to be clusters.  The two clusters (MS2137.3-2353 and
MS1512.4+3647) which were classed as ambiguous in \scite{nic97} are also
found to be clusters in HRI observations by \scite{mol99}, which adds
further doubt to the robustness of the classification scheme of \scite{nic97}.

The second way in which our reanalysis differs with that of
\scite{nic97} is in the data reduction method.  We have used a spectral
cube when performing our background subtraction.  This allows the photon
vignetting to be calculated more accurately as a function of energy
than when using an image in a single passband, as in \scite{nic97}, when a
mean photon energy must be assumed.

Finally, the third way in which our reanalysis is different to that of
\scite{nic97} is in the data analysis.  \scite{nic97} measure count rates within 
a fixed metric aperture that is calculated (as a function of redshift and
PSF) to encompass 85 \% of the total flux.  However this calculation is
made assuming a King profile of $\beta = \frac{2}{3}$ and a core radius of
$r_{\textrm{\scriptsize{c}}}=250$ kpc.  We improve upon this by measuring the core radius of each 
cluster and correcting from detected flux to total flux using the King
profile with the measured core radius (see \S2.2).  The count rates were
converted to fluxes assuming a 6 keV Raymond-Smith spectrum in \scite{nic97}, whereas we had
the advantage of being able to use measured temperatures (for all but one
cluster) from ASCA measurements.

\subsection{Flux measurement.}

For each cluster with a PSPC observation the data were reduced in the
following way.  The raw data were downloaded from the LEDAS ROSPUB archive.
The data were first cleaned for bad times and a corrected exposure time was
obtained.  A spectral cube was sorted, of $384 \times 384$ spatial bins and
10 energy bins, and a background cube was associated with it.  Cubes were
used as this makes correcting for photon vignetting more accurate than
if images were used when a mean photon energy would have to be assumed.  The
background was measured in an annulus from $0.15^{\circ}$ -- $0.25^{\circ}$
to avoid the original target of the pointing and the shadows of the window
support structure.  The cluster was masked out of this background annulus
if it were coincident with it.  Point sources within the background annulus
were removed. The point sources were found using the \emph{ASTERIX  PSS}
routine (\pcite{all95}) 
which employs  a maximum likelihood search based on the \scite{cas79} 
statistic.  The point
source searching procedure was iterated until no more significant point sources were
found.  A final check of the background was made by eye.  Any obvious
contaminations were removed and the procedure was re-run.  
The expected particle background was calulated using the method of \scite{sno92}.
The measured
value of the background was then extrapolated over the field as a function
of energy, taking  into account the unvignetted particle background, and subtracted 
from the source.  The
background-subtracted cube was then corrected for energy-dependent vignetting.  
The exposure was corrected for dead time.

From the background-subtracted cube the number of counts within an aperture
of 3Mpc in radius was extracted  (this corresponds to $9.09'$ at $z=0.3$
and $6.57'$ at $z=0.6$).  Point sources within the aperture found by inspection
of raw and smoothed images,
were masked out (except in the cluster cores, where the high surface brightness
cluster emission may have masked faint point sources).  
An aperture of 3 Mpc was chosen as the correction 
to total flux is small and it is much larger than the PSPC
psf even at an off-axis angle of $30'$.The average error on the count rate was 
$\sim 6\%$ based on
counting statistics.  

The counts within the aperture were 
corrected assuming a King profile of
$\beta = \frac{2}{3}$ and initially a core radius of $r_{\textrm{\scriptsize{c}}} = 250$kpc, to
obtain the total number of counts.  The correction factor for such a
profile is 1.09.  This correction is based on integrating the emission out
to infinity.  Whilst it may be more physically accurate to integrate out
only as far as the virial radius (\pcite{hen00}), our aim is to
compare our $ROSAT$ luminosities with the previous measurements of 
\scite{gio94} and \scite{nic97}, and to the local XLF of \scite{ebe97},
who all integrated to infinity.
\scite{hen00} notes that because of other effects in the EMSS, 
although the published EMSS luminosities were integrated to infinite radius, on average  
they agree with luminosities
integrated only to the virial radius, and thus may be correct.
We wish to investigate in detail these, and other, effects, 
in the EMSS luminosities. In order to do this, and to better understand  differences
with previous work, we will initially follow the
method of previous investigations and integrate to infinite radius,
but then discuss the effects of integrating only to the virial radius.

To convert from count rates to fluxes a \scite{ray77} spectral model 
was assumed.  A model spectrum was constructed for each cluster and
wherever possible the model parameters  were taken from the literature.
For the 
cluster MS0418.3-3844 no measured temperature exists and therefore 
$T_{\textrm{\scriptsize{X}}}$ was assumed to be 6 keV and for all clusters except MS0451.5-0305
the metallicity is assumed to be $0.3 \times$ solar.  The hydrogen column
densities were taken from \scite{dic90}.  Absorbed and unabsorbed fluxes and luminosities were
measured in the
appropriate energy bands.  The model takes into account
K-corrections when calculating luminosities.

Exceptions to the method described above were made in the cases of
MS0353.6-3642 and MS0418.3-3844.  Both these clusters fell very close to
the shadow of the window support structure in the PSPC image, and consequently extracting a
count rate over a 3 Mpc radius would not yield accurate results.  For these
clusters radial profiles were taken to estimate the radius at which the
surface brightness fell to the background level and the counts within that
radius were measured.  These radii were $0.04^{\circ}$ and $0.05^{\circ}$,
corresponding to 823 kpc and 1080 kpc, for  MS0353.6-3642 and MS0418.3-3844 
respectively.  The counts were corrected for the same King profile as
previously and converted to fluxes and luminosities in an identical manner
as before.

\subsection{Surface brightness fitting and determination of $r_{\textrm{\scriptsize{c}}}$.}

Measurements of the core radii, $r_{\textrm{\scriptsize{c}}}$, were made for each cluster to
allow a more accurate conversion from the detected count rate to the total
count rate.  This was carried out by fitting a King profile to the surface
brightness taking into account blurring by the PSPC psf.  The free parameters were the right--ascension and
declination of the cluster centre, the core radius, and the normalisation.  The
index, $\beta$ was frozen at $\frac{2}{3}$ and it was assumed that the
cluster was symmetric.
The measured value of the core radius was then used  to recompute the total flux.

\section{Results.}

\subsection{The new luminosities.}

A summary of the results of the recalculated luminosities is shown in
Table~1 and Table~2.

\begin{table*}
\caption{A comparison of luminosities as measured by various authors.
 \dag means the count rate was extracted over a small aperture and \ddag
  means the temperature of the cluster was assumed to be 6 keV.  $^{*}$ The 
  quoted EMSS luminosity for MS1241.5+1710 has been scaled from the
  \protect\scite{gio94} value according to the new redshift measurement.}

\begin{tabular}{|l|l|l|l|l|l|l|} 
        & & \multicolumn{4}{c|}{Luminosity $\times 10^{44}$ ergs
          s$^{-1}$} (0.3 -- 3.5 keV) & \\
 Name & Redshift & EMSS & Nichol & Ellis ($r_{\textrm{\scriptsize{c}}}=250$ kpc) & Ellis ($r_{\textrm{\scriptsize{c}}}$
 measured) & $r_{\textrm{\scriptsize{c}}}$ /kpc \\
 \hline
MS0015.9+1609 & 0.546 & 14.64$\pm 1.63$ & 22.78$\pm 0.54$ & 27.96$\pm 0.71$ & 27.93$\pm 0.71$ & $247^{+5}_{-8}$ \\ 
MS0353.6-3642\dag & 0.320 & 5.24$\pm 1.06$ & 8.70$\pm 0.73$ & 8.65$\pm 0.74$ & 8.33$\pm 0.71$ & $225^{+37}_{-31}$ \\ 
MS0418.3-3844\dag \ddag & 0.350 & 1.433$\pm 0.264$ & 1.49$\pm 0.096$ & 4.84$\pm 0.33$ & 6.61$\pm 0.43$ & $474^{+92}_{-75}$ \\ 
MS0451.6-0305 & 0.55 & 19.98$\pm 3.58$ & 22.36$\pm 1.14$ & 30.02$\pm 1.43$ & 29.16$\pm 1.39$ & $168^{+8}_{-7}$ \\ 
MS0811.6+6301 & 0.312 & 2.10$\pm 0.42$ & 1.99$\pm 0.5$ & 5.30$\pm 0.73$ & 5.35$\pm 0.73$ & $280^{+77}_{-61}$ \\ 
MS1241.5+1710 & 0.549 & 9.05$^{*} \pm 2.31$ & & 6.05$\pm 1.23$ & 5.85$\pm 1.19$ & $204^{+30}_{-27}$ \\ 
MS1358.4+6245 & 0.327 & 10.62$\pm 1.86$ & 9.65$\pm 0.29$ & 10.65$\pm 0.44$ & 10.22$\pm 0.43$ & $132^{+4}_{-5}$ \\ 
MS1426.4+0158 & 0.320 & 3.71$\pm 0.48$ & 5.38$\pm 0.46$ & 7.08$\pm 0.78$ & 7.21$\pm 0.75$ & $302^{+43}_{-38}$ \\ 
MS1512.4+3647 & 0.372 & 4.81$\pm 1.05$ & 5.83$\pm 0.54$ & 3.20$\pm 0.71$ & 3.05$\pm0.68$ & $114^{+14}_{-14}$ \\ 
MS1621.5+2640 & 0.426 & 4.55$\pm 0.856$ & & 10.35$\pm 1.05$ & 11.28$\pm 1.14$ & $480^{+95}_{-81}$ \\ 
MS2137.3-2353 & 0.313 & 15.62$\pm 2.13$ & 17.11$\pm 0.45$ & 19.03$\pm 0.62$ & 17.73$\pm 0.57$ & $49^{+2}_{-3}$ \\ \hline

        \end{tabular}

\end{table*}

\begin{table*}
\caption{The ratios of luminosities measured here to previous
  measurements.  The
  symbols \dag and \ddag are the same as in Table~1.  The
          symbol $\clubsuit$ refers to the averages when the clusters
          MS0418 and
          MS0353 are not included, for a correct comparison.}
\begin{tabular}{|l|l|l||l|l|} 
        Name & \multicolumn{2}{c||}{Ratio ($r_{\textrm{\scriptsize{c}}}=250$ kpc)} &
\multicolumn{2}{c|}{Ratio ($r_{\textrm{\scriptsize{c}}}$ measured)} \\ 
 & EMSS & Nichol & EMSS & Nichol \\ \hline
MS0015.9+1609 & 1.91$\pm 0.22$ & 1.23$\pm 0.04$ & 1.91$\pm 0.22$ & 1.23$\pm 0.04$ \\ 
MS0353.6-3642\dag & 1.65$\pm 0.36$ & 0.99$\pm 0.12$ & 1.59$\pm 0.35$ & 0.96$\pm 0.11$ \\ 
MS0418.3-3844\dag \ddag & 2.61$\pm 0.53$ & 2.51$\pm 0.28$ & 3.38$\pm 0.69$ & 3.25$\pm 0.36$ \\ 
MS0451.6-0305 & 1.50$\pm 0.29$ & 1.34$\pm 0.09$ & 1.46$\pm 0.27$ & 1.30$\pm 0.09$\\ 
MS0811.6+6301 & 2.52$\pm 0.61$ & 2.66$\pm 0.76$ & 2.55$\pm 0.62$ & 2.69$\pm 0.77$ \\ 
MS1241.5+1710 & 0.67$\pm 0.22$ & & 0.65$\pm 0.21$ & \\ 
MS1358.4+6245 & 1.00$\pm 0.18$ & 1.10$\pm 0.06$ & 0.96$\pm 0.17$ & 1.06$\pm 0.54$ \\ 
MS1426.4+0158 & 1.91$\pm 0.32$ & 1.32$\pm 0.18$ & 1.94$\pm 0.32$ & 1.34$\pm 0.18$ \\ 
MS1512.4+3647 & 0.66$\pm 0.21$ & 0.55$\pm 0.13$ & 0.63$\pm 0.20$ & 0.52$\pm 0.13$ \\ 
MS1621.5+2640 & 2.28$\pm 0.49$ & & 2.48$\pm 0.53$ & \\ 
MS2137.3-2353 & 1.22$\pm 0.17$ & 1.11$\pm 0.05$ & 1.14$\pm 0.16$ & 1.04$\pm 0.04$ \\  
Average & 1.28$\pm 0.08$ & 1.16$\pm 0.02$ & 1.22$\pm 0.07$ & 1.13$\pm 0.03$ \\ 
$\clubsuit$ & 1.23$\pm 0.08$ & 1.16$\pm 0.03$ & 1.18$\pm 0.08$ & 1.13$\pm 0.03$\\ \hline

        \end{tabular}

\end{table*}

The total luminosities are on (error-weighted)
average $1.28 \pm 0.08$ times greater than those in \scite{gio94} and
$1.16 
\pm 0.02$ times
greater than those in \scite{nic97} with $r_{\textrm{\scriptsize{c}}}=250$ kpc.  Using the
measured values of $r_{\textrm{\scriptsize{c}}}$ these values become $1.22 \pm 0.07$ and $1.13
\pm 0.03$
respectively.  A figure more representative of the actual average increase
in luminosity is found by excluding the clusters MS0353.6-3642
and MS0418.3-3844, for which the count rate was extracted over a small
aperture (and hence open to larger errors in correcting for a King
profile).  

Excluding these clusters the error-weighted average ratios become $1.23 \pm 0.08 \times$
EMSS and
$ 1.16 \pm 0.03 \times$ Nichol for $r_{\textrm{\scriptsize{c}}}=250$ kpc and $1.18 \pm 0.08
\times$ EMSS and
$1.13 \pm 0.03 \times$ Nichol for measured $r_{\textrm{\scriptsize{c}}}$.  This latter value of
1.18  $\times$ EMSS will be used to correct the luminosities of the EMSS
clusters for which there were no PSPC data available.  Throughout the rest
of this paper all comparisons will be made using luminosities calculated using
measured core radii.

Note that the ratio for MS1241.5+1710 is calculated using the luminosity
scaled from that given in \scite{gio94} according to the new redshift measurement.

Four clusters in the sample have been reanalysed following a similar
aperture method by \scite{jon98a}.  We find that our fluxes in this paper are 
an average factor of $1.12 \pm 0.02$ times larger than those measured by
\scite{jon98a}.  Although the detailed analysis used here is an improvement over  
that used by \scite{jon98a}, since we include core radii measurements,
this comparison gives an estimate of the
absolute accuracy of our luminosities.

\subsubsection{Luminosity differences as a function of core radius.}

An important point is that the ratio of $ROSAT$ aperture luminosities
to EMSS luminosities is a relatively strong function of measured core radii
(see Fig. 3).
The error-weighted average ratio for clusters with
$r_{\textrm{\scriptsize{c}}}<250$kpc, $1.10 \pm 0.03$ (68\% confidence error on the mean)
is significantly lower than that for
clusters with $r_{\textrm{\scriptsize{c}}}>250$kpc, $2.30 \pm 0.12$ (or $1.08 \pm 0.03$  for $r_{\textrm{\scriptsize{c}}}<250$kpc
and $2.16 \pm 0.15$ for $r_{\textrm{\scriptsize{c}}}>250$kpc excluding MS0353.6-3642
and MS0418.3-3844). The correlation of $L_{\textrm{\scriptsize{ap}}}/L_{\textrm{\scriptsize{EMSS}}}$ with $r_{\textrm{\scriptsize{c}}}$ is 
significant at $>$99.9\% confidence (correlation coefficient of 0.85).
This correlation contributes part of the observed scatter on the mean luminosity
ratio of all the clusters (1.18$\pm$0.08).

The absolute broad-band  flux calibration accuracy of the ROSAT PSPC
 ($\approx$10\%; \pcite{bri96}, \emph{ROSAT} users handbook) and
of the Einstein IPC (also $\approx$10\%; \pcite{har84}, \emph{EINSTEIN} revised
user's manual) are smaller than the differences
found, for example, in Fig 3, although they probably contribute part of the 
systematic difference observed.  We note, however, that any calibration
uncertainties cannot explain the observed correlation of
$\frac{L_{\textrm{\scriptsize{ap}}}}{L_{\textrm{\scriptsize{EMSS}}}}$ with core radius.

A key point made in this paper is that while the clusters with small core radii
show reasonable agreement between $ROSAT$ large aperture luminosities and EMSS
luminosities, within the calibration uncertainties,
the clusters with large core radii have significantly higher 
$ROSAT$ luminosities.

\subsection{Reasons for the luminosity differences.}

To investigate possible causes for the differences between our results and
those of the EMSS  we show plots of $\frac{L_{\textrm{\scriptsize{ap}}}}{L_{\textrm{\scriptsize{EMSS}}}}$ against
$L_{\textrm{\scriptsize{ap}}}$, $z$ and $r_{\textrm{\scriptsize{c}}}$ for clusters at $z>0.3$ in Fig.~1, Fig.~2 and Fig.~3. 
The errors in the ratios
are calculated from counting statistics in the PSPC data combined with errors in
the EMSS luminosities.

\begin{figure}
\psfig{file=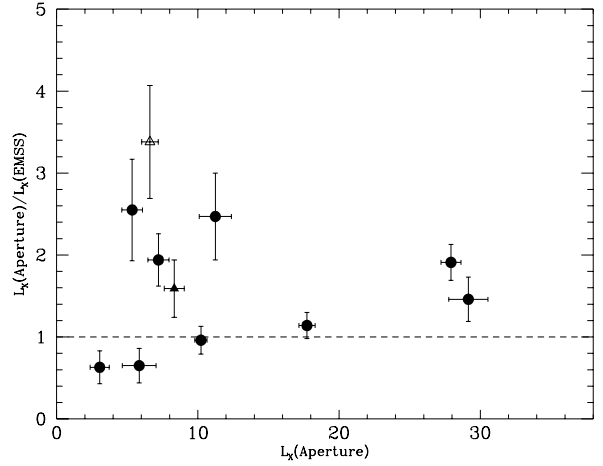,width=8.5cm,height=6.5cm,angle=270}
\caption{$\frac{L_{\textrm{\scriptsize{ap}}}}{L_{\textrm{\scriptsize{EMSS}}}}$ as a function of $L_{\textrm{\scriptsize{ap}}}$ (in units of
 $10^{44}$ erg s$^{-1}$).  Circles
  denote that the counts were measured over a 3 Mpc radius,  and triangles
  over a smaller radius.  Open symbols denote that the temperature was assumed to 
  be 6 keV, whilst closed symbols denote that the data have measured
  temperatures.}
\end{figure}

\begin{figure}
\psfig{file=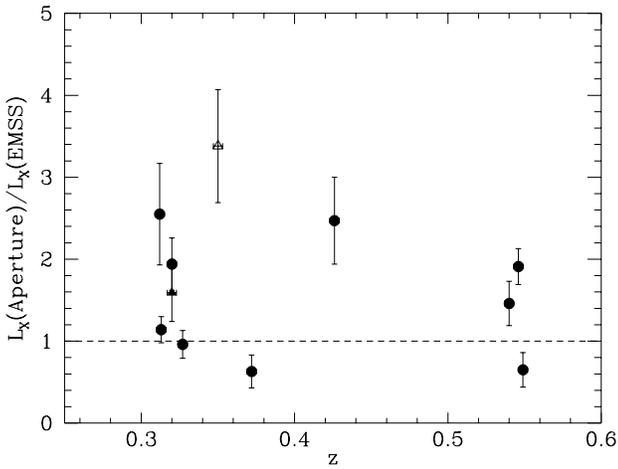,width=8.5cm,height=6.5cm,angle=270}
\caption{$\frac{L_{\textrm{\scriptsize{ap}}}}{L_{\textrm{\scriptsize{EMSS}}}}$ as a function of  $z$.  Symbols are the
  same as in Fig~1.}
\end{figure}

\begin{figure}
\psfig{file=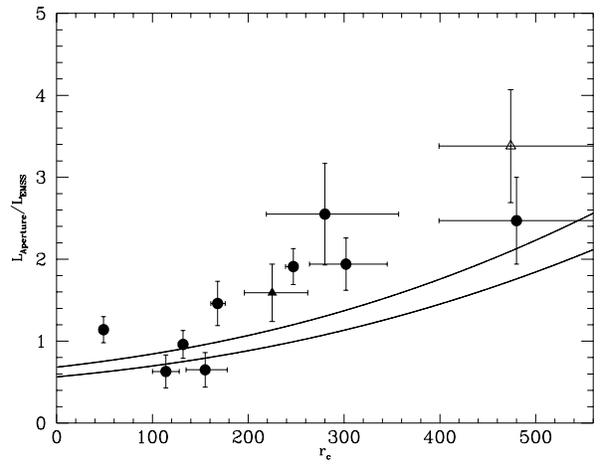,width=8.5cm,height=6.5cm,angle=270}
\caption{$\frac{L_{\textrm{\scriptsize{ap}}}}{L_{\textrm{\scriptsize{EMSS}}}}$ as a function of core radius (in units of 
  h$_{50}^{-1}$ kpc).  Symbols are the
  same as in Fig~1.  The lower line is the predicted ratio of the true
  luminosity to that assuming $r_{\textrm{\scriptsize{c}}}$ = 250 kpc; the upper line is the the
  lower one multiplied by the correction for the IPC psf blurring (a factor 
  of 1.21).  See the discussion for details. }
\end{figure}

Figs. 1 and 2 show that there is no obvious trend of luminosity ratio
with luminosity or redshift.  
Note that the highest point representing MS0418.3-3844 may
be considered as misleading for reasons discussed above.  
The luminosity of the cluster
MS0015.9+1609 given by the EMSS may be too low as it lies at the extreme edge of
the IPC field (\pcite{hen00}).   

The reason for the difference in luminosity for clusters of high core radius 
may have its origins in the conversion of 
detected count rate to total flux.  There are two stages in this
conversion when an error could be introduced to the conversion factor.
These are  the conversion from
detected count rate to total count rate and the conversion from total count rate
to flux.

The conversion from count rate to flux is dependent on the spectral model
used and the parameters specific to it.  We have used a Raymond and Smith
spectral model following \scite{hen92}, but we have included previously
unavailable measured values for the X-ray temperature, $T_{\textrm{\scriptsize{X}}}$.  To examine
the size of the effect
that the value of $T_{\textrm{\scriptsize{X}}}$ used in the spectral model has on the
conversion factor the flux and luminosity of the coolest (MS1512.4+3647)
and hottest (MS0451.6-0305) cluster were recalculated assuming $kT = 6$ keV 
as in \scite{hen92}.  For MS1512.4+3647 $\frac{f(6 keV)}{f(3.57 keV)} =
1.12$ and $\frac{L(6 keV)}{L(3.57 keV)} = 1.06$, and for MS0451.6-0305 $\frac{f(6 keV)}{f(10.4 keV)} =
0.94$ and $\frac{L(6 keV)}{L(10.4 keV)} = 1.00$.  Thus it can be seen
that the conversion from count rate to flux is relatively insensitive to
the exact value of temperature used in the spectral model and the
discrepancy must have its origins in the conversion from detected to total count rate.

A possible source of error in the conversion from detected to total count rate is 
the assumed surface brightness profile used.  This is especially important
when converting from the detected count rate
in the $2.4' \times
2.4'$ detect cell of the EMSS to the total count rate as
the  correction is large.  This correction is very sensitive to the
model parameters used, especially the value used for the core radius.  In
\scite{hen92} it was assumed that $r_{\textrm{\scriptsize{c}}} = 250$ kpc, whereas our surface
brightness fitting gave measured values of $r_{\textrm{\scriptsize{c}}}$ ranging from 49 kpc to
480 kpc (albeit with rather large errors on the larger clusters).  It would 
be expected that for clusters whose actual core radius is less than 250 kpc 
the EMSS would \emph{overestimate} the conversion from detected to total
count rate. Conversely,  clusters with $r_{\textrm{\scriptsize{c}}} > 250$ kpc would be  underestimated 
in the EMSS as there would be more emission lying outside the detect cell than accounted for 
assuming  $r_{\textrm{\scriptsize{c}}} = 250$ kpc. 

Fig~3 shows that 
there is a trend where $\frac{L_{\textrm{\scriptsize{ap}}}}{L_{\textrm{\scriptsize{EMSS}}}}$ increases with
$r_{\textrm{\scriptsize{c}}}$, as expected. 
The underestimation becomes progressively bigger for larger 
clusters.  
The lower solid line in Fig. 3 gives
the predicted luminosity ratio based on the constant $r_{\textrm{\scriptsize{c}}}=250$kpc of \scite{hen92}, ie.
the predicted ratio is unity for $r_{\textrm{\scriptsize{c}}}=250$kpc. Most clusters lie above this line.
Thus the EMSS assumption of a constant core radius cannot explain all of
the observed discrepancy. There may be a further systematic difference between
the luminosities.

\subsection{The \emph{EINSTEIN} IPC PSF.}

Another source of possible error has been pointed out by \scite{hen00};
the conversion from detected flux to total flux in \scite{hen92}
(equation 1 of that paper) does not include the effect of the IPC point spread
function (psf). Furthermore \scite{hen00} calculates that the effect of including the
psf would be to  increase the factor used to convert the number of counts in the 
detect cell to total counts, $1/f$, by 1.373 for clusters
at $0.3<z<0.6$. This effect would explain the
discrepancy we have measured.

The effect of the IPC psf was investigated further.  Firstly it was checked 
whether the psf  of the IPC is dependent on the off-axis angle of the
source.  Twenty four bright AGN in the EMSS survey (\pcite{gio90}) were
selected and a Gaussian profile was fitted to a radial plot of surface
brightness in the EMSS detection energy band of 0.2 -- 3.5 keV.  It was found that the full width at half maximum of the Gaussian did not show
any trends with off-axis angle.  The reason for this lack of trend is that
the electronics of the IPC dominate the scattering from the mirrors (\pcite{gia79}).  Having established that the psf of the IPC+mirror
combination is independent of off-axis angle, the on-axis psf was used to
investigate what effect this would have on the fraction of measured flux to 
total flux.  A pure Gaussian is used to model the IPC psf since the flux
lost into the non-Gaussian wings of the combined IPC+mirror psf, at large
radii, caused by mirror scattering, has already been accounted for in the
EMSS fluxes.  The psf used was taken from the average measurement of the
AGN Gaussian profiles and had FWHM$=92'' \pm 5''$.

For each cluster a model King profile was constructed from the measured values of 
core radii and an assumed value of $\beta = \frac{2}{3}$.  This was then
blurred by convolving it with the IPC psf.  From this dataset the ratio of
counts, $f_{\textrm{\scriptsize{psf}}}$, falling within a $2.4' \times 2.4'$ square centred on the cluster
and a circle of radius $30'$ (approximately the total counts) was measured.  This was compared to the
predicted ratio between the same two apertures assuming a pure King profile 
without blurring, $f_{\textrm{\scriptsize{King}}}$, as used in \scite{hen92}.

For the eleven clusters remeasured in this work the assumption of a pure
King profile gives total counts that are on average a factor of 0.83 too
small, i.e. the original EMSS fluxes need increasing by a factor of 1.21.  This is
slightly lower than, but still comparable to, the result obtained in \scite{hen00} of an
increase of 1.373 in converting from counts in a $2.4' \times 2.4'$ cell to
total counts.  The difference in the two results is due mainly to the size
of the psf used in calculating $f_{\textrm{\scriptsize{psf}}}$ (\pcite{hen00} uses FWHM$=105''$,
private communication) and partly due to the inclusion of 
measured core radii of individual clusters.  As a check we calculated
$f_{\textrm{\scriptsize{psf}}}$ for an assumed 250 kpc cluster in the redshift range $z=0.3$--0.6 
and find values of $\frac{f_{\textrm{\scriptsize{King}}}}{f_{\textrm{\scriptsize{psf}}}}$ of 1.28 and 1.34 for psfs of
$92''$ and $105''$ respectively.  We have elected to use the value 
of FWHM=$92''$ as measured from IPC data (we note that the FWHM for
clusters, with harder IPC spectra than AGN, may if anything be less than $92''$).

The redshift dependence of the effect of the psf on the conversion from
detected to total counts is shown in figure~4.  This was calculated for an
on-axis cluster of $r_{\textrm{\scriptsize{c}}} = 250$ kpc.  It can be seen that between $z=0.15$
and $z=0.6$ there is almost no change in $\frac{f_{\textrm{\scriptsize{psf}}}}{f_{\textrm{\scriptsize{King}}}}$; this is
because the psf is larger than the cluster at such high redshifts.  

\begin{figure}
\psfig{file=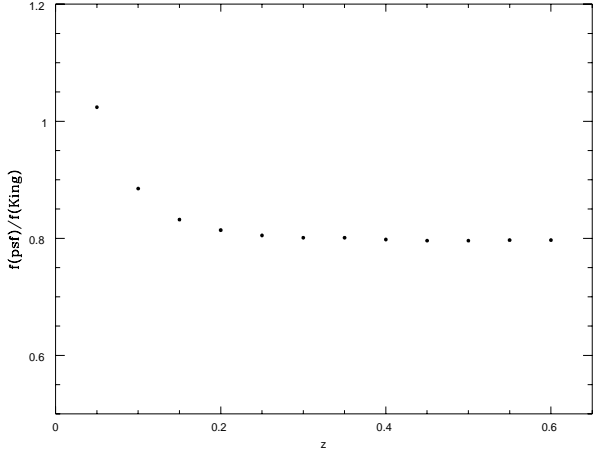,width=8.5cm,height=6.5cm,angle=270}
\caption{$\frac{f_{\textrm{\scriptsize{psf}}}}{f_{\textrm{\scriptsize{King}}}}$ as a function of redshift.}
\end{figure}


\subsection{The new X-Ray luminosity function.}

Using the new luminosities discussed above the X-ray luminosity function
was computed and compared to the EMSS XLF computed using the data 
from \scite{gio94}.  In the sample 
of \scite{gio94} there are 24 clusters with $z > 0.3$ and measured
luminosities.  In our sample taken from \scite{gio94} there are only 21
clusters having these criteria.  It is only these 21 clusters which are
used in computing the luminosity functions.  (The reasons for the
differences between the samples are that MS1333.3+1725 has now been
identified as a star (\pcite{lup99}), MS1209.0+3917 as a BL Lac
(\pcite{rec98}) and MS1610.4+6616 as a point source at the position of a
star (\pcite{sto99}).  Note that MS0354.6-3650 which appears in
\scite{hen92} does not appear in \scite{gio94} and appears to be a soft
X-ray source (\pcite{nic97}).)

The XLF was computed using the $\frac{1}{V_{a}}$ method of \scite{avn80} as 
used in the non-parametric analysis of \scite{hen92}.  This method is
summarised here.

For each cluster the luminosity distance and the angular size of the core
radius (as measured) were calculated according to the formulae below, 

\begin{equation}
D_{L}=\frac{c}{q_{0}^{2}H_{0}}(q_{0}z + (q_{0}-1)(\sqrt{1+2q_{0}z}-1)) 
\end{equation}
\begin{equation}
\theta_{0}=\frac{r_{\textrm{\scriptsize{c}}}(1+z)^2}{D_{L}}
\end{equation}

The fraction of counts which fell inside the detect cell of $2.4' \times 2.4'$ was calculated according to,

\begin{equation}
f=\frac{2}{\pi} \arcsin
\left\{
\frac{(\frac{\theta_{D}}{\theta_{0}})^2}{((\frac{\theta_{D}}{\theta_{0}})^2+1)} 
\right\}
(g_{\textrm{\scriptsize{psf}}})
\end{equation}

where $\theta_{D}$ is the angular half size of the detect cell and $g_{\textrm{\scriptsize{psf}}} = \frac{f_{\textrm{\scriptsize{psf}}}}{f_{\textrm{\scriptsize{King}}}}$ is a new factor which takes into account the effect of the
IPC psf.  As fig~4 shows, the effect of the IPC psf is constant for a particular
cluster above $z=0.3$, so the inclusion of $g_{\textrm{\scriptsize{psf}}}$ is unimportant here as it
will cancel out in equation~4 below.

The maximum redshift at which the cluster could have been
detected was calculated for each flux limited observation making up the
survey.  This was found by incrementing $z_{max}$ in the following formula until the statistic, $FIT$ gave the value closest to 1, where

\begin{equation} 
FIT=\frac{F_{DET}}{F_{LIM}} \left( \frac{D_{L}(z)}{D_{L}(z_{max})} \right)^{2}
  \left( \frac{f(z_{max})}{f(z)} \right)
\end{equation}

where $F_{DET}$ is the detect cell flux at the observed redshift $z_{obs}$, 
and $F_{LIM}$ is the limiting survey flux.

The total volume in which each cluster could have been found was then
calculated by summing over all the flux limits 
\begin{eqnarray}
V_{a} & = & \sum_{i} \left[ \frac{dV(\Omega_{0},\leq
    min(z_{u},z_{max,i}))}{d\Omega} \right. \nonumber \\
& & \left. - \frac{dV(\Omega_{0},\leq z_{l})}{d\Omega} \right] d\Omega_{surv,i}  
\end{eqnarray}

where $z_{u}$ and $z_{l}$ are the upper and lower redshift limits of the
sample and $d\Omega_{surv,i}$ is the solid angle associated with the $ith$
flux limit.  The mean ratio of the search volumes found here to those
assuming no blurring from the IPC psf and a constant core radius of $r_{\textrm{\scriptsize{c}}} = 
250$kpc (as in \scite{hen92}) was $0.99 \pm 0.03$.

The values for $\frac{dV}{d\Omega}$ were calculated by numerical
integration of the following equation
\begin{equation}
\frac{dV}{d\Omega dz} = 4 \left( \frac{c}{H_{0}} \right)^{3}
\frac{[\Omega_{0}z+(2-\Omega_{0})(1-\sqrt{1+\Omega_{0}
    z})]^{2}} {\Omega_{0}^{4}(1+z)^{3} \sqrt{1+\Omega_{0} z}}
\end{equation}
with respect to $z$.

The clusters were binned up into log luminosity bins that are 0.3 wide and
then the differential luminosity function was calculated for each bin
\begin{equation}
n(L)=\sum_{j=1}^n \frac {1}{V_{a,j} \Delta L}
\end{equation}
where $\Delta L$ is the width of the luminosity bin and $n$ is the number
of clusters in that bin.

The XLF is recomputed using the remaining 21 clusters from \scite{gio94}
and compared to our XLF in two ways.  Firstly using our new
luminosities for those clusters observed by the PSPC and the luminosities
of \scite{gio94} for those clusters not observed by the PSPC.  This is shown 
in Fig.~5.  Secondly the luminosities of the clusters not observed by the
PSPC are corrected by the average increase measured (when not including
MS0418.3-3844 and MS0353.6-3642).  This is shown in Fig.~6.

\begin{figure}
\psfig{file=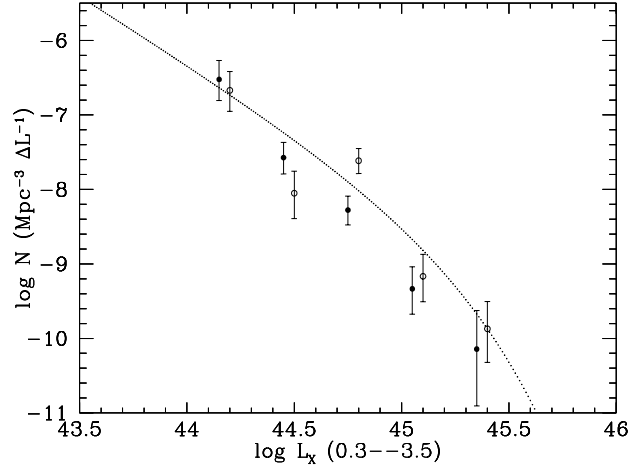,width=8.5cm,height=6.5cm,angle=270}
\caption{The revised X-ray Luminosity Function at $z = 0.3$ --- 0.6.  Closed circles are from the data of
  \protect\scite{gio94} and open circles are from PSPC data.  The dotted
line is
  the local BCS XLF from \protect\scite{ebe97}.}
\end{figure}

\begin{figure}
\psfig{file=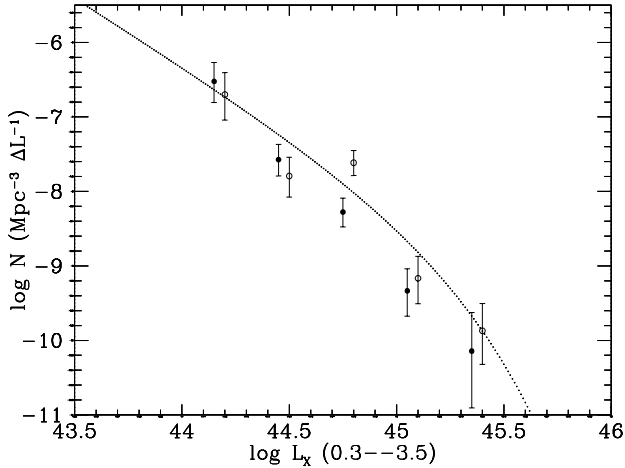,width=8.5cm,height=6.5cm,angle=270}
\caption{The revised X-ray Luminosity Function at $z$ = 0.3 --- 0.6.  Symbols are the same as in Fig.~5
  except that clusters for which there are no PSPC data have had their
  luminosities corrected by the average correction factor of 1.18.}
\end{figure}

\begin{figure}
\psfig{file=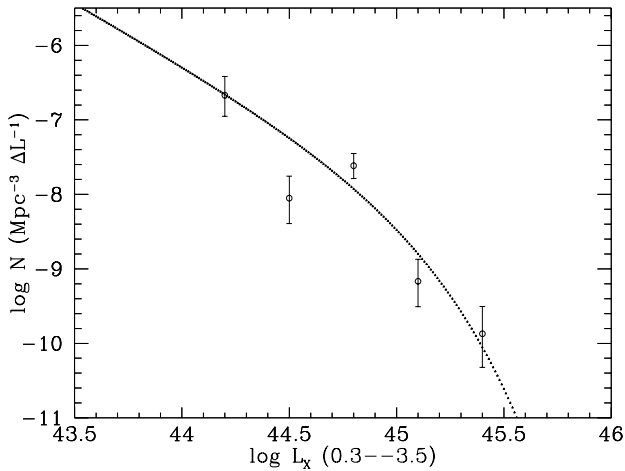,width=8.5cm,height=6.5cm,angle=270}
\caption{The revised X-ray Luminosity Function at $z$ = 0.3 --- 0.6, compared 
  to the local XLF of \protect\scite{boh01}. The data points are based 
  on the revised luminosities, including an average correction factor for 
  clusters with no PSPC data, but all luminosities have then been decreased 
  by 8\%, to be consistent with the approximation used by B{\"o}hringer et al. for
  integrating only to the virial radius. } 
\end{figure}

Superimposed on both Fig.~5 and Fig.~6 is the best fitting Schechter
function to the local XLF from \scite{ebe97} which is obtained from the
\emph{ROSAT} Brightest Cluster Sample (BCS). All the luminosities 
we have used  to calculate the XLFs in Figs. 5 and 6 are based on an integration of the
emission to infinity, as is the BCS XLF. 
In Fig. 7 we compare the updated XLF with the local XLF 
of \scite{boh01}, which updates the work of \scite{deg99} and which was published while this paper was being produced. 
The  B{\"o}hringer et al. XLF   was also converted from the
0.1--2.4 keV band to the 0.3--3.5 keV band. For cluster temperatures of $T_{\textrm{\scriptsize{X}}}=4$--$10$ keV,
a conversion factor of $L_{0.3-3.5}=1.1 L_{0.1-2.4}$ is accurate to within
$\approx$8\%. B{\"o}hringer et al. use luminosities measured by integrating to the 
virial radius, approximated by 12 times the core radius. As noted by \scite{boh00}, this approximation gives a systematic decrease of 8\% compared 
to integrating to infinity. 
The same data as in Fig. 6 are plotted, except that all the luminosities
were reduced by 8\%, to be consistent with the virial radius  approximation 
used by B{\"o}hringer et al..

It is clear that whereas
previously there existed evidence for a degree of evolution from the local XLF the
effect of the increase in luminosity is to bring the high
redshift XLF back in line with it.  This appears to be the case in 
Figs.~5, 6 and 7.


\subsection{Tests of evolution.}

An alternative test for the presence of evolution
is to compare the observed number of clusters at 0.3$<z<$0.6 to the number predicted
based on a non-evolving local XLF and the sky coverage of the survey.

Two local XLFs were used in this test, that of \scite{ebe97} and that of \scite{boh01}.
The local XLF was integrated over redshift (0.3$<z<$0.6) and over luminosity
($10^{43}-10^{46}$ erg s$^{-1}$), to predict the number of clusters as a function 
of total flux. K-corrections were included for the 0.3-3.5 keV band
based on temperatures estimated from the $L_{\textrm{\scriptsize{X}}}$--$T$ relation of \scite{whi97}, and calculated from
\emph{MEKAL} model spectra. The total fluxes were then converted to detect-cell
fluxes using a single value of $f$, and the EMSS sky coverage of \scite{hen92}
(which is defined as a function of  detect-cell flux)
used to give the total number of clusters predicted in the EMSS.

For the local XLF of \scite{ebe97}, the published 0.3-3.5 keV XLF could be
used directly. \scite{ebe97} used luminosities calculated by integrating 
the surface brightness to infinite radius, so the appropriate values of $f$ 
to use were 
$f_{\textrm{\scriptsize{psf}}}=0.488\pm0.05$, the mean value of the 11 clusters studied here 
including the effect of their measured core radii, and
$f_{\textrm{\scriptsize{psf}}}=0.445$, the value for a cluster with $r_c=250$kpc at z=0.387
(the mean redshift of the EMSS 0.3$<$z$<$0.6 sample).
For comparison, $f_{\textrm{\scriptsize{King}}}=0.593\pm0.06$, again the mean over the 11  
clusters, was also used. Using  $f_{\textrm{\scriptsize{King}}}$ neglected the effect of the 
\emph{EINSTEIN} IPC psf. 

The local XLF of \scite{boh01} has a different normalisation definition 
to that of \scite{ebe97}, and was also converted from the
0.1-2.4 keV band to the 0.3-3.5 keV band as above.  B{\"o}hringer et al. only
integrate the emission to the virial radius.
The approximation to the virial radius 
used by B{\"o}hringer et al.   gives an decrease of 8\% compared 
to integrating to infinity, independently  of the value of the core radius.
Thus the values of $f$ we use are the values of $f_{\textrm{\scriptsize{psf}}}$ as used for
the  \scite{ebe97} XLF, but multiplied by 1.08. 

The results are shown in Table 3.  The number of clusters predicted (N$_{pred}$)
assuming no evolution of the XLF is given in the third column. In all
cases this is more than the number observed ($N_{\textrm{\scriptsize{obs}}}=21\pm4.6$). 
The mean of the two best no-evolution predictions is 42, twice the number
observed. The significance
of the difference, in units of sigma, is given in the following column. 
Errors on the predicted numbers were derived from the square root of 
the total number of clusters in each of 
the local surveys.

\begin{table*}
\caption{Results of the tests of the evolution of the XLF.}
\begin{minipage}{12cm}
\begin{tabular}{|l|l|l||l|l|l|} 
        
Local XLF & $f$ & $\frac{1}{f}$ & $N_{\textrm{\scriptsize{pred}}}$ & Significance of & Notes on $f$ \footnote{See text for full explanation} \\ 
 & & & & $N_{\textrm{\scriptsize{pred}}} - N_{\textrm{\scriptsize{obs}}}$
 \footnote{$N_{\textrm{\scriptsize{obs}}} = 21 \pm 4.6$}
 ($\sigma$) & \\ \hline 
BCS \footnote{BCS XLF of \scite{ebe97}} & 0.488 & 2.05 & $35.9 \pm 2.5$ & 2.8 & $\overline{f_{\textrm{\scriptsize{psf}}}}$; mean over 11 clusters\\ 
BCS & 0.445 & 2.25 & 31.3 $\pm 2.2$ & 2.0 & $f_{\textrm{\scriptsize{psf}}}$
for $r_{\textrm{\scriptsize{c}}}=250$ kpc, $z=0.39$  \\
 
(BCS & 0.593 & 1.69 & 47.9 $\pm3.4$ & 4.7) \footnote{Use of
  $f_{\textrm{\scriptsize{King}}}$ rather than
  $f_{\textrm{\scriptsize{psf}}}$ is incorrect but is included to show the sensitivity of $N_{\textrm{\scriptsize{pred}}}$ on $f$.} & $\overline{f_{\textrm{\scriptsize{King}}}}$; mean over 11 clusters\\ 
\\
REFLEX \footnote{REFLEX XLF of \scite{boh01}} & 0.527 & 1.90 & 46.4 $\pm 2.2$ & 4.9 &
$\overline{f_{\textrm{\scriptsize{psf}}}} \times 1.08$ to account for the REFLEX \\
& & & & & luminosities within $12 r_{\textrm{\scriptsize{c}}}$ \\
REFLEX & 0.489 & 2.08 & 40.3 $\pm 1.9$ & 3.9 &
$f_{\textrm{\scriptsize{psf}}} \times 1.08$; for
$r_{\textrm{\scriptsize{c}}}=250$kpc, $z=0.39$ \\
REFLEX & 0.427 & 2.34 & 33.1 $\pm 1.6$ & 2.5 &
$f_{\textrm{\scriptsize{psf}}} \times 1.08$; for
$r_{\textrm{\scriptsize{c}}}=300$kpc, $z=0.39$ \\
REFLEX & 0.545 & 1.83 & 49.0 $\pm 2.3$ & 5.5 &
$f_{\textrm{\scriptsize{psf}}} \times 1.08$; for
$r_{\textrm{\scriptsize{c}}}=200$kpc, $z=0.39$ \\
(REFLEX & 0.640 & 1.56 & 62.5 $\pm 2.9$ & 7.6)$^{d}$
& $\overline{f_{\textrm{\scriptsize{King}}}} \times 1.08$\\\hline

        \end{tabular}

\end{minipage}
\end{table*}

\section{Discussion.}

\subsection{The new luminosities.}

The ratio of the luminosities of clusters with $z > 0.3$ as measured by our 
aperture photometry using \emph{ROSAT} PSPC data and as
measured in the original EMSS has been investigated. The ratio is found to be 
correlated with the core radius of the cluster. 
For clusters with small core radii the \emph{ROSAT} and EMSS
luminosities are in agreement, within the instrumental calibration uncertainties.   
For clusters with progressively larger core radii, the \emph{ROSAT}
luminosities increase to $\approx$2 times the EMSS luminosities.
A straightforward explanation of the correlation is 
that for larger clusters ($r_{\textrm{\scriptsize{c}}} > 250$ kpc) more flux
would fall outside the detect cell than predicted by \scite{hen92}, who
assumed a constant 250kpc core radius.  

An additional difference is probably caused by the omission of the \emph{Einstein} IPC
point spread function when converting from detected flux to total flux, as
suggested by \scite{hen00}. 
We have shown that the inclusion of the psf would increase the EMSS fluxes
by a factor $\approx 1.21$ for clusters at $z > 0.3$.  The combination of this factor together with
the measured core radii can account for most of the observed difference in
luminosity as shown in Fig~3.  The lower line in Fig~3 shows the predicted
ratio of true luminosity to the luminosity derived from an EMSS detect cell 
flux, assuming a fixed core radius of 250kpc (at a redshift of z = 0.4).
The upper line shows the effect of including IPC psf blurring in this prediction.
For clusters with small core radii ($<$250kpc), the overestimate of EMSS luminosity
due to the assumption of a fixed core radius may in part have been cancelled 
by the underestimate due to the effect of the \emph{Einstein} IPC psf. 
For clusters with large core radii ($>$250kpc), however, the two effects work
in the same direction, which would produce an underestimated EMSS luminosity.

\subsubsection{Integrating to the viral radius.}

In order to make comparisons with previous work, 
the luminosities discussed so far have  been derived using the same method as 
in those works, ie.
an extrapolation of a King profile to infinite radius (except when comparing
with the XLF of \pcite{boh01}). If instead we define the
luminosity $L_{\textrm{\scriptsize{vir}}}$, perhaps more accurately, as that within a King
profile truncated at the virial radius $r_{\textrm{\scriptsize{vir}}}$ (as suggested by \pcite{hen00}), where
$r_{\textrm{\scriptsize{vir}}} \approx 1.8$Mpc for a $T=6$keV cluster at $z=0.4$, then the
luminosities we calculate would decrease.  The size of the decrease depends on the core
radius and the temperature, but would typically be $\approx 13\%$ for a
$T=6$keV, $r_{\textrm{\scriptsize{c}}}=250$kpc cluster, and $3\%$--$25\%$ for
$r_{\textrm{\scriptsize{c}}}=50$--$500$kpc.  Thus, as noted by \scite{hen00}, the original EMSS
$z>0.3$ luminosities fortuitously give approximately accurate mean values of
$L_{\textrm{\scriptsize{vir}}}$, since the 13\% decrease in the $ROSAT$ luminosities would nearly
cancel the average 18\% difference between the $ROSAT$ and EMSS luminosities.

What would  $L_{\textrm{\scriptsize{vir}}}$ be for clusters with different core radii?
Using $r_{\textrm{\scriptsize{vir}}}=r_{200}=
3.89(T_{\textrm{\scriptsize{x}}}/10 $keV$)^{1/2} (1+z)^{-3/2}$ Mpc (for H$_0$=50; \pcite{evr96}), $r_{\textrm{\scriptsize{vir}}}=1.84$ Mpc for  a $T_{\textrm{\scriptsize{X}}}=6$keV cluster at $z=0.39$, the mean 
redshift of the 21 clusters in the EMSS sample.  For a core radius of 150 kpc,
the $ROSAT$ values of $L_{\textrm{\scriptsize{vir}}}$ would be 8\% lower than the luminosities  
quoted in Tables 1 and 2. The ratio of $ROSAT$ to EMSS luminosities for
clusters with  $r_{\textrm{\scriptsize{c}}}<250$kpc would then
be in even better agreement, with a mean ratio of 1.0$\pm$0.3. 

For a core radius of 350kpc, the $ROSAT$ values of $L_{\textrm{\scriptsize{vir}}}$ would be 
19\% lower than the luminosities  
quoted in Tables 1 and 2. The ratio of $ROSAT$ to EMSS luminosities for
clusters with  $r_{\textrm{\scriptsize{c}}}>250$kpc would then be 1.81$\pm$0.15, still
significantly larger than unity.   

Summarizing, for small ($r_{\textrm{\scriptsize{c}}}<250$kpc) clusters there is good agreement
between the $ROSAT$ and EMSS luminosities, especially if the $ROSAT$ 
luminosities are derived from an integration only within the virial radius.
For large clusters  ($r_{\textrm{\scriptsize{c}}}>250$kpc), the agreement is less good, and
here we suspect that the EMSS luminosities need increasing by a factor
of 1.8-2.2, depending on the radial integration limit used.  
How many clusters have large core radii?
In the local
sample of \scite{vik99} half of all clusters (19 of 39) 
have $r_{\textrm{\scriptsize{c}}}>250$kpc (consistent with the 4 out of the 11 clusters studied here), 
and a third have  $r_{\textrm{\scriptsize{c}}}>300$kpc. Thus a substantial fraction of the cluster
population is affected.

\subsubsection{$ASCA$ luminosities.}

It is also noted in \scite{hen00} that there is a discrepancy between
\emph{ASCA} fluxes and EMSS fluxes with ASCA fluxes being a factor 1.17 higher 
(or 1.12 excluding
MS0015.9+1609), a figure very consistent with the value 1.18$\pm 0.08$
found in this paper.  It is suggested in \scite{hen00} that the reason here is the contamination
of the measured flux by unrelated sources in the GIS extraction beam.  This 
explanation is in conflict with the previous explanation of the
\emph{ROSAT}/EMSS discrepancy.  The \emph{ROSAT}/EMSS discrepancy is
explained if the EMSS fluxes are \emph{increased} by 1.21 (the effect of
including the IPC psf), 
whereas the \emph{ASCA}/EMSS discrepancy is explained if the \emph{ASCA}
fluxes are \emph{decreased} to bring them into line with the 
EMSS fluxes (because of unrelated sources in the \emph{ASCA} beam).
Clearly both of the explanations cannot be true.




\subsection{Evolution of the cluster space density.}

We have made two tests of the degree of evolution based on  the EMSS 
cluster luminosities at  0.3$<z<$0.6. 

The comparison of the number of observed clusters with that predicted
from an integration of the local XLF over redshift and luminosity, assuming
no evolution, has some advantages. 
It includes the effect of non-detections, and does not use the luminosities directly,
only the total number of clusters above the flux limits of the survey. 
A disadvantage is that, for the EMSS, a correction has to be made  
from the total flux predicted (based on the
 integration of the local XLF),  to a detect
 cell flux. Without accurate knowledge of the distribution of surface 
brightness shapes at high redshift (eg. core radii), a mean value of the 
core radius has to be used for this correction. 

The results of the comparison are that for the BCS XLF of \scite{ebe97},
the observed number of clusters is a factor of 1.7 lower than the
no-evolution prediction, at a significance of 2.8$\sigma$.
For the REFLEX XLF of \scite{boh01}, the observed
number of clusters is a factor of 2.2 lower than the
no-evolution prediction, at a significance of 4.9$\sigma$.
These predictions are sensitive to the mean value of $f$ assumed;
for values of $f$ corresponding to mean core radii of 200 kpc -- 300 kpc,
the significance of the difference from no-evolution varies from 
5.5$\sigma$ to 2.5$\sigma$.

The second test was the updated binned EMSS XLF. 
The XLF has the  advantage of an individual correction to total
flux for each cluster (at least for those with measured core radii),
but has the disadvantages of excluding luminosities where no
clusters were detected, and the somewhat arbitrary choice of binning.
The updated XLF is consistent with no evolution, and, given the 
size of the error bars on the binned XLF, consistent with a factor 
of $\approx$2 decrease in the normalisation of the XLF at z$\approx$0.4
suggested by the first test.

When predicting the number of clusters from the local XLFs, we have used
Schechter function paramaterizations of the local XLFs. These are good fits
for most luminosities. However, at the highest luminosities ($L_{\textrm{\scriptsize{X}}}>10^{45}$
erg s$^{-1}$, 0.1-2.4 keV) Figs. 1 and 4 of \scite{boh01} show that
the Schechter function fit overestimates the number of clusters in the
REFLEX XLF by 
a factor $\approx$1.7 which increases with luminosity (the Schechter function fit
underestimates the number of clusters at lower luminosities
 $L_{\textrm{\scriptsize{X}}}\approx10^{44}$ erg s$^{-1}$, but by a smaller
factor $\approx$1.3).  The cause of the overestimate at high luminosities may be
small number statistics or real deviations from a Schechter function.
In either case, if the number of high luminosity clusters observed in the
REFLEX survey were used, rather than the Schechter function fit, 
the effect would be to reduce the no-evolution prediction. 
Approximately 25\% of the
EMSS sample have $L_{\textrm{\scriptsize{X}}}>10^{45}$ erg s$^{-1}$, 
and $\approx$15\% have  $L_{\textrm{\scriptsize{X}}}<2$x$10^{44}$ erg s$^{-1}$, 
so this would be 
a relatively small net effect, reducing the predictions in Table 3 by $\approx$10\%,
and therefore reducing the 4.9$\sigma$ significance to 4.2$\sigma$.
The significance would be reduced slightly further if we based the errors in the
predicted numbers on only the number of clusters in the local XLFs that have
luminosities which match those in the 0.3$<$z$<$0.6 EMSS sample 
ie $L_{\textrm{\scriptsize{X}}}>10^{44}$ erg s$^{-1}$, rather than the full 
number in the local XLF.

Another uncertainty arises from the suggestion of
\scite{ebe00} that the EMSS may be incomplete due to a systematic bias against
unrelaxed clusters.  Ebeling et al. suggest that the bias may be  more pronounced for high luminosity
clusters at intermediate to high redshift, providing an alternative
explanation for some of the negative evolution observed in the EMSS.

In summary, based on one of two local XLFs, we find that the predicted number of 
clusters may be inconsistent with no evolution (but with caveats),
whilst using another local XLF (but containing a smaller number of
clusters) gives a less significant result. Both no-evolution predictions are 
a factor of $\approx$2 higher than the observed number of clusters, although
the predictions are sensitive to $f$, the correction from total to
detect cell fluxes, and we have used a mean value of $f$.
The binned XLF is consistent with no evolution, but also with 
a factor of $\approx$2 evolution in the number of clusters. Overall,
we conclude that the space density of 0.3$<z<$0.6 EMSS clusters,
based on their updated luminosities, is consistent with either no
evolution or a factor of $\approx$2 fewer clusters.

A factor of two evolution in the cluster space density 
is consistent with that found by \scite{hen92}
at $L_{\textrm{\scriptsize{X}}}=3 \times 10^{44}$ erg s$^{-1}$, but 
we do not find larger evolutionary factors at higher luminosities,
as suggested by an extrapolation of the power law XLF fits of \scite{hen92}.
At the highest luminosities sampled, 
$L_{\textrm{\scriptsize{X}}}=1 \times 10^{45}$--$3 \times 10^{45}$ erg s$^{-1}$
(0.3--3.5 keV), there are 5$^{+3.4}_{-2.2}$ clusters in the updated sample,
compared with 9.1 predicted from a non-evolving REFLEX XLF (using $f_{psf}$=0.527),
and 9.0 predicted from the BCS XLF (using $f_{psf}$=0.488). Evolution by a large
factor of 9 in the XLF normalisation, for example, giving a prediction
of one cluster at these luminosities, is ruled out at $>$99\% confidence by the 5 observed.


The little or no  evolution of the XLF observed in this work has important consequences 
for future and past work
that concern cluster evolution based on the EMSS sample.  As was stressed
in the introduction the EMSS is very suited to constraining cluster
evolution due to the (relatively) high number of high luminosity,
massive clusters at high redshift which it contains. 

One measurement which can be made using
the constraint of little or no evolution of the XLF is to constrain the matter density
parameter of the Universe, $\Omega_{\textrm{\scriptsize{0}}}$.  Little or no  evolution favours a
low $\Omega_{\textrm{\scriptsize{0}}}$ universe (\pcite{kay99}).  This result is in contrast with that 
of previous determinations of
$\Omega_{\textrm{\scriptsize{0}}}$ based on the apparent evolution in the EMSS XLF.
\scite{rei99} find $\Omega_{\textrm{\scriptsize{0}}} = 0.96$ from their analysis of the evolution
of the high redshift sample of EMSS XLF compared to the BCS.  \scite{sad98}
also derive a high value of $\Omega_{\textrm{\scriptsize{0}}}$ ($=0.85 \pm 0.2$) from their work
on the $L_{\textrm{\scriptsize{X}}}$--$T_{\textrm{\scriptsize{X}}}$ relation and the redshift distribution of  the EMSS
sample which is supported by the work of \scite{bla98}.   

\subsection{Search volumes and consequences for the X-ray temperature function.}

The revised  luminosities of clusters in the EMSS could
also have an effect on other studies based on the EMSS sample.
The new core radii not only change the values of the
luminosity used in calculating the XLF but they also slightly alter the search
volumes used.  Because the new volumes are not systematically larger or 
smaller than search volumes
calculated assuming $r_{\textrm{\scriptsize{c}}} =250$ kpc and no blurring due to the IPC psf they 
only have a small effect on the XLF.  The
neglect of the IPC psf blurring has almost no effect on the calculated volumes,
because $g_{\textrm{\scriptsize{psf}}}$ cancels out in equation~4 and $g_{\textrm{\scriptsize{psf}}}$ is essentially
independent of redshift at $z>0.3$.

\scite{hen00} and \scite{don99} use
temperatures measured from \emph{ASCA} for the EMSS high-z sample to
determine the evolution of the temperature function, and derive important measurements
of $\Omega_0$. Search volumes are determined in a similar way as described above,
from values of $f$ as a function of redshift, and detect cell fluxes.
We find that the search volumes are affected little by the assumption of 
a core radius of 250 kpc compared to using the measured core radii 
(the difference for the clusters with $r_c>250$ kpc
is $\leq$3\%), or by neglecting the IPC psf blurring (since the effect of  
 $g_{\textrm{\scriptsize{psf}}}$ cancels out). 


\section{Conclusions.}

The X-ray luminosities of 11 EMSS clusters at 0.3$<$z$<$0.6 have been remeasured
using  \emph{ROSAT} PSPC data, including a measurement of the core radii,
and using large apertures to avoid uncertain extrapolations of the surface
brightness.  
For clusters with core radii $r_c<250$ kpc, we find reasonable agreement between 
the \emph{ROSAT} luminosities and the 
 original luminosities of \scite{hen92} and \scite{gio94} (a mean ratio of 1.08$\pm$0.03). For 
clusters with large core radii, $r_c>250$ kpc, the \emph{ROSAT} luminosities
are 2.2$\pm$0.15 times the EMSS luminosities.  
This difference can be largely explained by the 
assumption of a fixed core
radius (of 250 kpc) in the EMSS, combined with 
the omission of the effect of the \emph{EINSTEIN} IPC psf in
calculating the total fluxes. 
These effects are most important in deriving accurate luminosities. 
The effect on the EMSS cluster survey
search volumes, which are used in both luminosity function and temperature function 
studies, is small.

The binned EMSS XLF at 0.3$<$z$<$0.6, based on the new luminosities and on an updated sample, is consistent
with no evolution. The predicted numbers of clusters, assuming no evolution and based on
two local luminosity functions, is $\approx$2 times the number observed at 0.3$<$z$<$0.6. 
Given the uncertainties
in the luminosities, in the derivation of detect cell fluxes,
and the relatively small numbers of luminous clusters in both the EMSS and
in the local samples,  we conclude that the updated EMSS sample is consistent with either
no evolution of the space density of luminous clusters, or a small factor of $\approx$2 
fewer luminous clusters at z=0.4 (the mean redshift of the sample). 
Little or no evolution in the space denisty of luminous clusters is consistent
with low values of  $\Omega_{0}$.

\section{Acknowledgments.}

The authors would like to thank Pat Henry for useful discussions.  SCE
acknowledges a PPARC studentship; LRJ also acknowledges PPARC support.  This research has made
use of data obtained from the Leicester Database and Archive Service at the
Department of Physics and Astronomy, Leicester University, UK and the
Einline service at the Harvard-Smithsonian Center for Astrophysics.  The
authors would also like to thank the referee for some important improvements.

\bibliographystyle{mnras}
\bibliography{clusters}

\end{document}